\documentclass[a4paper,11pt]{article}
\usepackage{pos}
\usepackage[capitalise]{cleveref}
\def\Dz{D^0\to K^-\pi^+}
\def\Dp{D^+\to K^-\pi^+\pi^+}
\def\pDz{D^0}
\def\pDp{D^+}
\def\Lbc{\Lambda_c^+\to pK^-\pi^+}
\def\pLbc{\Lambda_c^+}
\def\Omgc{\Omega_c^0\to\Omega^-\pi^+}
\def\pOmgc{\Omega_c^0}
\title{Measurements of charm lifetimes at Belle II}
\author*[1]{N.~K.~Nisar }

\affiliation[a]{Brookhaven National Laboratory,\\
  Upton, New York, 11973}

\note{On behalf of the Belle II Collaboration}
\emailAdd{nnellikun@bnl.gov}

\abstract{We report on absolute lifetime measurements of charmed hadrons using the data collected by the Belle II experiment between 2019 and 2021. The measured lifetimes of $D^0$, $D^+$, and $\Lambda_c^+$ are the most precise to date and consistent with previous measurements. Our result indicates that $\Omega_c^0$ is not the shortest-living singly charmed baryon.}

\FullConference{%
  41st International Conference on High Energy physics - ICHEP2022\\
  6-13 July, 2022\\
  Bologna, Italy
}

\begin{document}
\maketitle

\section{Introduction}

Predictions of beauty and charm hadron lifetimes are achieved by the heavy quark expansion (HQE) model~\cite{Neubert:1997gu,Uraltsev:2000qw,Lenz:2013aua,Lenz:2014jha,Kirk:2017juj,Cheng:2018rkz}.
The charm lifetime predictions are particularly challenging due to the significant higher-order corrections and spectator quark effects.
So the charm lifetime measurements allow for HQE validation and refinement that increase the reliability and precision of Standard Model predictions in flavor dynamics. The best measurements of charm meson lifetimes date back to FOCUS~\cite{FOCUS} while LHCb recently reported precise measurements of charm baryon lifetimes, relative to $D^+$ lifetime~\cite{Ll_prd,LHCb:omgc,LHCb:omgc2}.

We report absolute lifetime measurements of the charm hadrons using the data collected by the Belle II detector~\cite{belle2}, which is built around the interaction region (IR) of the SuperKEKB~\cite{skekb} asymmetric energy $e^+e^-$ collider. SuperKEKB adopts a nano-beam scheme that squeezes the IR to achieve large instantaneous luminosity. The Belle II detector consists of a tracking system, a particle identification system, and an electromagnetic calorimeter kept inside a 1.5 T superconducting magnet. The outer layer consists of a dedicated muon and $K_L^0$ detector. The details of the Belle II detector can be found in Ref.~\cite{belle2}. Excellent vertex resolution, precise alignment of the vertex detector, and accurate calibration of particle momenta in Belle II are crucial in the measurements of lifetimes.       

\section{Lifetime extraction}

The proper decay times of charm hadrons are calculated as $t=m(\vec{L}\cdot\hat{p})/p$, where $m$ is the known mass of hadrons, $\vec{L}$ is the flight length between the production and decay vertices, and $p$ is the momentum of hadrons. Lifetimes are extracted by using unbinned maximum-likelihood fits to the $t$ and its uncertainty, $\sigma_t$, of the candidates populating the signal regions of data. The signal probability-density function (PDF) is the convolution of an exponential function in $t$ with a resolution function that depends on $\sigma_t$, multiplied by the PDF of $\sigma_t$. The time constant of the exponential function will return the lifetime. The $\sigma_t$ PDF is a histogram template derived directly from the signal region of the data.
In all cases but $D^0$, the template is obtained from the candidates in the signal region after having subtracted the distribution of the sideband data. Simulation demonstrates that for $\pDp$, $\pLbc$, and $\pOmgc$, a single Gaussian function is sufficient, whereas for $\pDz$, a double Gaussian function with a common mean is required.

\section{$D^0$ and $D^+$ lifetimes}
\begin{figure}[t!]
\centering
\includegraphics[width=0.5\linewidth]{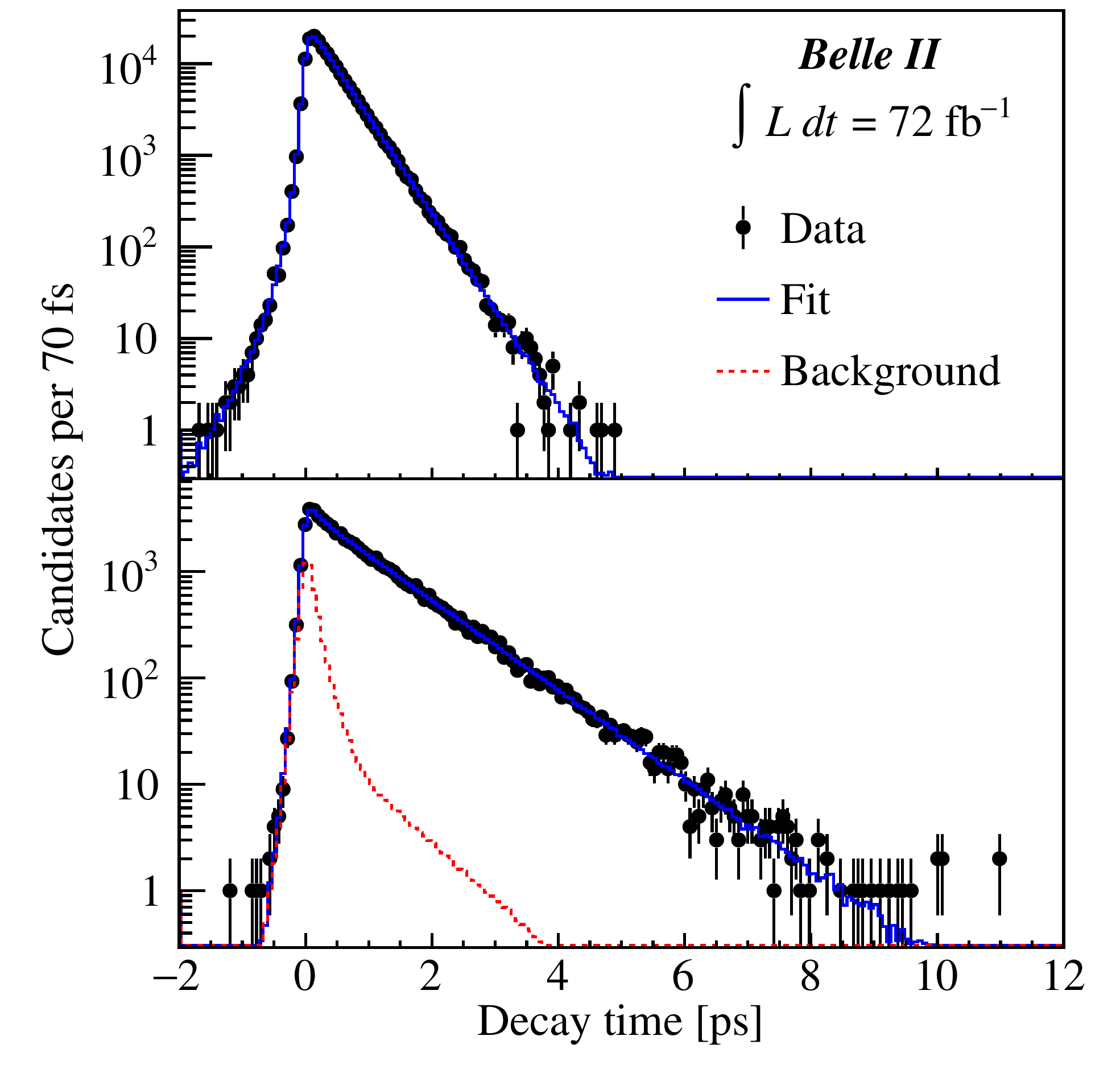}\\
\caption{Decay-time distributions of (top) $\Dz$ and (bottom) $\Dp$ candidates in their respective signal regions with fit projections overlaid. \label{fig:lifetime_D}}
\end{figure}

We measured $\pDz$ and $\pDp$ lifetimes using $\rm72~fb^{-1}$ of Belle II data using samples of reconstructed $\Dz$ and $\Dp$ decays, respectively. 
$171\times10^3$ signal candidates are reconstructed for $D^{*+}\to\pDz(\to K^-\pi^+)\pi^+$ decays in the signal region: $1.851<m(K^-\pi^+)<1.878~{\rm GeV}/c^2$. 
In the $D^0$ case, the per-mille-level fraction of background candidates in the signal region is neglected, and a systematic uncertainty is assigned for this. $59\times10^3$ signal candidates are reconstructed for $D^{*+}\to\pDp(\to K^-\pi^+\pi^+)\pi^0$ decays in the signal region: $1.855<m(K^-\pi^+\pi^+)<1.883~{\rm GeV}/c^2$.
For the $D^+$ case, a sizable background contamination in the signal region is accounted for using the data sideband: $1.758<m(K^-\pi^+\pi^+)<1.814~{\rm GeV}/c^2, 1.936<m(K^-\pi^+\pi^+)<1.992~{\rm GeV}/c^2$. The background PDF consists of a zero-lifetime component and two exponential components, all convolved with the resolution function.  The decay-time distributions of the data, with fit projections overlaid, are shown in Fig.~\ref{fig:lifetime_D}. The $D^0$ and $D^+$ lifetimes are measured to be $410.5\pm 1.1\pm0.8$~fs and $1030.4\pm4.7\pm3.1$~fs, respectively~\cite{dl_prl}. The errors are statistical and systematic (all relevant effects are studied as summarized in \cref{tab:D}), respectively. The results are consistent with their respective world average values~\cite{pdg}.

\begin{table}[t]
\centering
\caption{Systematic uncertainties for $\pDz$ and $\pDp$ lifetimes.\label{tab:D}}
\begin{tabular}{lcc}
\hline
Source & $\tau(\Dz)$ [fs] & $\tau(\Dp)$ [fs]\\
\hline
Resolution model   & 0.16 & 0.39 \\
Backgrounds        & 0.24 & 2.52 \\
Detector alignment & 0.72 & 1.70 \\
Momentum scale     & 0.19 & 0.48 \\
\hline
Total  & 0.80 & 3.10 \\
\hline
\end{tabular}
\end{table}

\section{$\Lambda_c^+$ lifetime}
The most precise measurement of the $\pLbc$ lifetime is reported by the LHCb experiment~\cite{Ll_prd}. We report a preliminary  result on the absolute measurement of the $\pLbc$ lifetime in $\Lbc$ decays reconstructed using $207~\rm fb^{-1}$ of the Belle II data. We reconstruct $116\times10^3$ candidates for the decay $\Lbc$ in the signal region: $2.283<m(pK^-\pi^+)<2.290~{\rm GeV}/c^2$, with a background contamination of 7.5\%. The $\pLbc$ lifetime is extracted in the same way as the $\pDp$ lifetime. Background events in the signal region are constrained using data sideband ($2.249<m(pK^-\pi^+)<2.264~{\rm GeV}/c^2$, $2.309<m(pK^-\pi^+) <2.324~{\rm GeV}/c^2$). 

Decays of $\Xi_c^0\to\pi^-\pLbc$ and $\Xi_c^+\to\pi^0\pLbc$ may bias
the measurement of the $\pLbc$ lifetime, since the $\Xi_c^0$ and $\Xi_c^+$ have non-zero lifetimes and may shift the production vertex of the $\pLbc$ away from the IR. A veto is applied to suppress such candidates, and a systematic uncertainty is assigned for the remaining contamination (details can be found in Ref.~\cite{Ll_prl}).  We measure the $\pLbc$ lifetime to be $\rm203.20\pm0.89\pm0.77~fs$, where the uncertainties are statistical and systematic (summarized in the \cref{tab:Lbc}), respectively~\cite{Ll_prl}. Our result is consistent with the current world average~\cite{pdg}.

\begin{figure}[t!]
\centering
\includegraphics[width=0.5\linewidth]{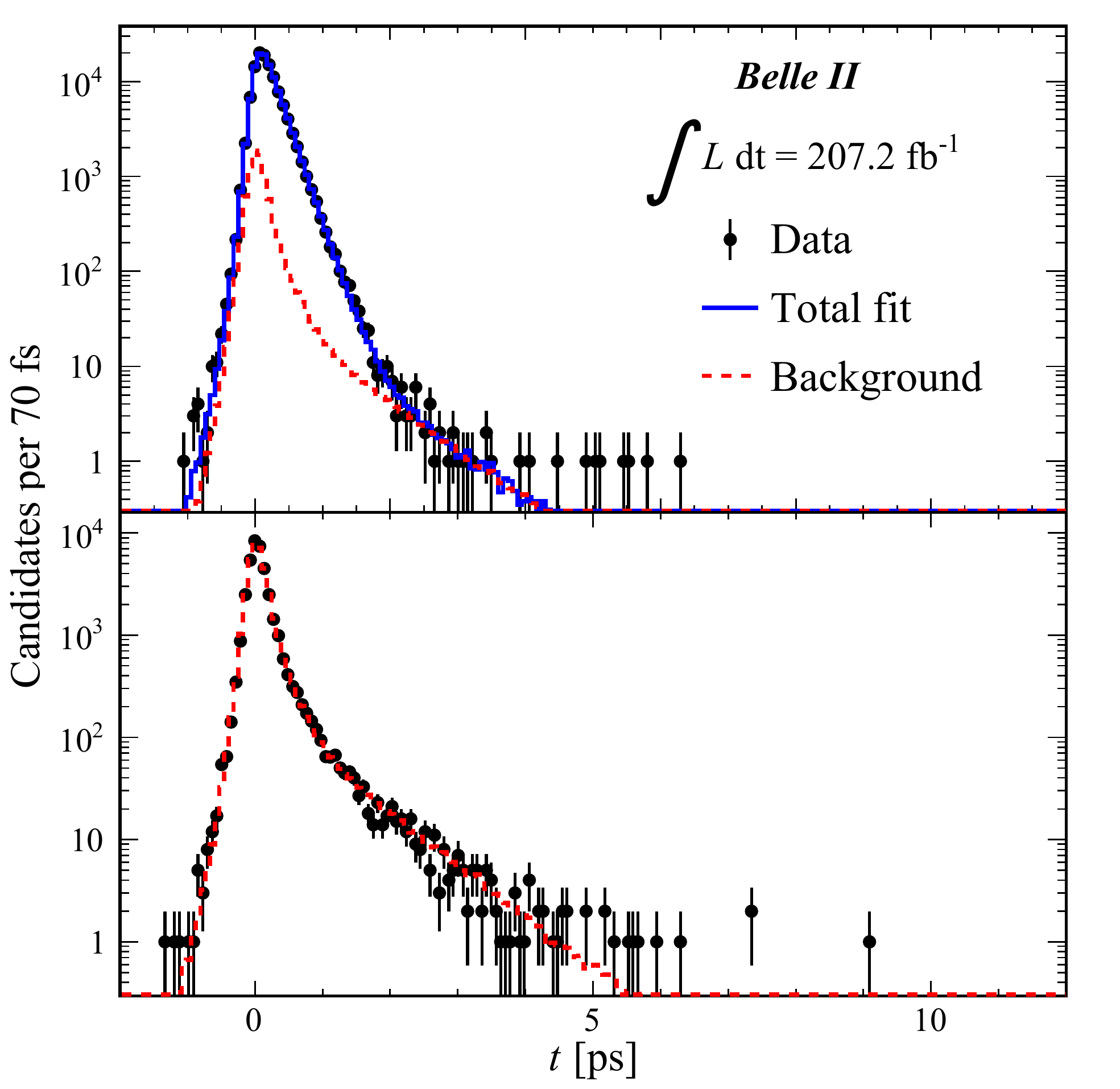}\\
\caption{Decay-time distributions of $\Lbc$ candidates in their (top) signal and (bottom) sideband regions with fit projections overlaid. \label{fig:lifetime_Lbc}}
\end{figure}

\begin{table}[t]
\centering
\caption{Systematic uncertainties for $\pLbc$ lifetime.\label{tab:Lbc}}
\begin{tabular}{lc}
\hline
Source & Uncertainty (fs) \\
\hline
$\Xi_c$ contamination  &  0.34 \\
Resolution model   &  0.46 \\
Non-$\Xi_c$ background model   &  0.20 \\
Detector alignment &  0.46 \\
Momentum scale     &  0.09 \\
\hline
Total & 0.77  \\
\hline
\end{tabular}
\end{table}

\section{$\Omega_c^0$ lifetime}
The $\Omega_c^0$ was believed to be the shortest-living singly charmed baryon that decays weakly. In 2018, LHCb measured a large value of $\Omega_c^0$ lifetime~\cite{LHCb:omgc}, and this observation inverted the lifetime hierarchy of singly charmed baryons. LHCb confirmed their result in 2022 using a different data sample~\cite{LHCb:omgc2}. We performed the first independent measurement of $\pOmgc$ lifetime using $\rm207~fb^{-1}$ of data collected at Belle II. We reconstructed 90 signal candidates in the signal region ($2.68<m(\Omega^-\pi^+)<2.71~{\rm GeV}/c^2$) for the decay $\Omgc$, where $\Omega^-\to\Lambda^0(\to p\pi^-) K^-$.  It is a complex decay chain with two extra decay vertices in addition to the $\Omega_c^0$ decay vertex.  
The lifetime is extracted by fitting the signal and sideband regions simultaneously. The signal region has a background contamination of 33\% that is constrained using events in the sideband ($2.55<m(\Omega^-\pi^+)<2.65~{\rm GeV}/c^2$, $2.75<m(\Omega^-\pi^+)<2.85~{\rm GeV}/c^2$). 
The $\pOmgc$ lifetime is measured to be  $\rm243\pm48\pm11~fs$, where the uncertainties are statistical and systematic (summarized in~\cref{tab:syst_omgc}), respectively~\cite{Ol_prl}. The result is consistent with LHCb measurements and inconsistent with previous measurements at 3.4 standard deviations.

\begin{figure}[t!]
\centering
\includegraphics[width=0.5\linewidth]{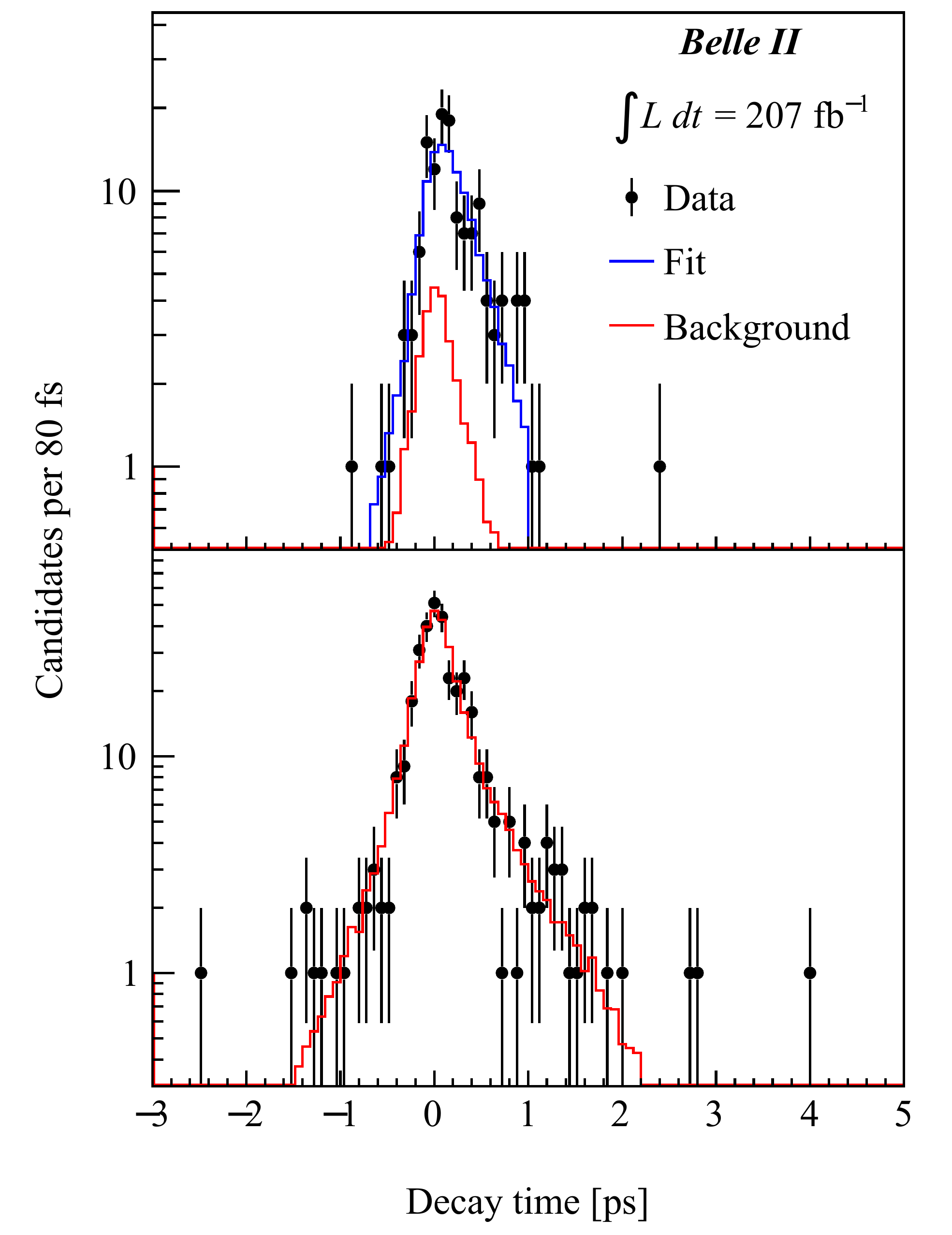}\\
\caption{
Decay-time distributions of $\Omgc$ candidates in their (top) signal and (bottom) sideband regions with fit projections overlaid.
\label{fig:lifetime_Omgc}}
\end{figure}

\begin{table}[t]
\centering
\caption{Systematic uncertainties for $\pOmgc$ lifetime.\label{tab:syst_omgc}}
\begin{tabular}{lc}
\hline
Source & Uncertainty (fs) \\
\hline
Fit bias           &  3.4 \\
Resolution model   &  6.2 \\
Background model   &  8.3 \\
Detector alignment &  1.6 \\
Momentum scale     &  0.2 \\
Input $\pOmgc$ mass   &  0.2 \\
\hline
Total & 11.0  \\
\hline
\end{tabular}
\end{table}

\section{Conclusions}
In conclusion, $\pDz$, $\pDp$, $\pLbc$, and $\pOmgc$ lifetimes are measured using the data collected by the Belle II experiment. The results on $\pDz$, $\pDp$, and $\pLbc$ lifetimes are the most precise to date and are consistent with previous measurements. Our result on $\pOmgc$ lifetime is consistent with the LHCb results~\cite{LHCb:omgc, LHCb:omgc2}, and inconsistent at 3.4 standard deviations with the pre-LHCb world average~\cite{pdg2018}. The Belle II result, therefore, confirms that the $\pOmgc$ is not the shortest-living weakly decaying charmed baryon.

\end{document}